\documentclass[prb,amsmath,amssymb,twocolumn,superscriptaddress,longbibliography]{revtex4-1}

\usepackage[dvipdfmx]{graphicx}
\usepackage{dcolumn}% Align table columns on decimal point
\usepackage{bm}% bold math
\usepackage{physics}% bold math
\usepackage{color}

\usepackage{amsmath}	% required for `\align' (yatex added)
\usepackage{amsfonts}
\usepackage{braket}
\usepackage{amsthm}

\makeatletter
\let\Caption\@makecaption
\makeatother
\usepackage{subcaption}
\makeatletter
\let\@makecaption\Caption
\makeatother

\begin{document}

%\preprint{APS/123-QED}

\title{Spin nematic order and superconductivity in $J_1$-$J_2$ Kondo lattice model
on square lattice}% Force line breaks with \\

\author{Sho Uraga}
\affiliation{Quantum Matter Program, Graduate School of Advanced Science and Engineering, Hiroshima University,
Higashihiroshima, Hiroshima 739-8530, Japan}

\author{Yasuhiro Tada}
\email[]{ytada@hiroshima-u.ac.jp}
%\homepage[]{Your web page}
%\thanks{}
\affiliation{Quantum Matter Program, Graduate School of Advanced Science and Engineering, Hiroshima University,
Higashihiroshima, Hiroshima 739-8530, Japan}
\affiliation{Institute for Solid State Physics, University of Tokyo, Kashiwa 277-8581, Japan}

%\date{\today}% It is always \today, today,
            %  but any date may be explicitly specified

\begin{abstract}
%\input{contents/Abst.tex}
% Abstract about this paper
We investigate competition and cooperation of magnetic frustration and the Kondo effect in the
$J_1$-$J_2$ Kondo lattice model on the square lattice at zero temperature.
In this model, the frustrated interactions $J_1,J_2$ between the localized spins stabilize spin nematic orders,
while the Kondo coupling favors local spin singlets.
Using the slave boson mean field approximation, 
we find that the spin nematic order remains stable against small Kondo coupling, and the localized spins
and the conduction electrons are effectively decoupled.
On the other hand, a standard Fermi liquid state is formed for sufficiently strong Kondo interactions.
Furthermore, in an intermediate region with moderate Kondo coupling, 
the spin nematic order and the Kondo effect coexist,
and superconducting pairing of the conduction electrons is induced by the spinon pairing.
We discuss the ground state phase diagram and nature of the quantum phase transitions between the different
superconducting states.
\end{abstract}

%\keywords{Suggested keywords}%Use show keys class option if keyword
                              %display desired
\maketitle

%\tableofcontents

\newcommand{\cdia}{c^{\dag}_{i \alpha}}
\newcommand{\cdib}{c^{\dag}_{i \beta}}
\newcommand{\cdja}{c^{\dag}_{j \alpha}}
\newcommand{\cdjb}{c^{\dag}_{j \beta}}
\newcommand{\cia}{c_{i \alpha}}
\newcommand{\cib}{c_{i \beta}}
\newcommand{\cja}{c_{j \alpha}}
\newcommand{\cjb}{c_{j \beta}}
\newcommand{\fdia}{f^{\dag}_{i \alpha}}
\newcommand{\fdib}{f^{\dag}_{i \beta}}
\newcommand{\fdja}{f^{\dag}_{j \alpha}}
\newcommand{\fdjb}{f^{\dag}_{j \beta}}
\newcommand{\fia}{f_{i \alpha}}
\newcommand{\fib}{f_{i \beta}}
\newcommand{\fja}{f_{j \alpha}}
\newcommand{\fjb}{f_{j \beta}}
\newcommand{\fie}{f_{i \eta}}
\newcommand{\fdjg}{f^{\dag}_{j \gamma}}
\newcommand{\fjg}{f_{j \gamma}}

%%%%%%%%%%%%%%%%%%%%%%%%%%%%%%%%%%%%%%%%%%%%%%%%%%%%%%%%%%%%%%%%%%
\section{\label{sec:level1}INTRODUCTION}

The competition and cooperation of a magnetic order and the Kondo effect 
are a fundamental subject in correlated electron systems
and have been extensively studied for many decades.
In the standard Kondo lattice, 
the Ruderman-Kittel-Kasuya-Yosida (RKKY) interaction typically favoring an antiferromagnetic order
and the Kondo screening effect giving rise to spin singlet formation compete each other.
Basic properties of the standard Kondo lattice model are summarized in the well known phase diagrams
~\cite{doniach1977kondo, Tsunetsugu1997}.

However, when there is magnetic frustration, 
the antiferromagnetic state may become unstable and
the phase diagram will be qualitatively modified~\cite{Senthil2003,coleman2010frustration,Vojta2018}. 
If the frustration is sufficiently strong, a conventional magnetic order can be completely suppressed and a spin liquid 
with fractionalized excitations could be realized in the localized spin sector.
When the spin liquid is gapped,
the localized spins and the conduction electrons will remain effectively decoupled even in prensece of non-zero 
Kondo coupling.
This is in sharp contrast to the conventional scenario where any non-zero Kondo interaction leads to 
a magnetic order of the whole system or a paramagnetic heavy Fermi liquid (HFL) whose Fermi surface consists of
both the conduction electrons and the localized spins~\cite{Luttinger1960,Oshikawa2000}.
On the other hand, in an effectively decoupled state, the spin liquid keeps its fractionalized nature and
the Fermi surface contains only the conduction electrons, which is beyond the standard Luttinger's theorem on the 
Fermi surface~\cite{Senthil2003,coleman2010frustration,Vojta2018}.
The resulting state is called the fractionalized Fermi liquid (FL$^*$) and it is stable for small Kondo coupling
and becomes unstable toward the standard HFL at sufficiently large Kondo coupling.

Interestingly, there can be intermediate phases between the FL$^*$ and HFL, and superconductivity
can be induced by an unconventional spin order
~\cite{Senthil2003,coleman2010frustration,Vojta2018,Seifert2018,choi2018topological,Bernhard2011,Pixley2014,su2015dimer}.
The emergent superconductivity has been studied for generalized Kondo lattice models with various 
magnetic interactions.
For example,
in a gapped $\mathbb{Z}_2$ spin liquid system, 
a superconducting phase can be found in between the FL$^*$ and the HFL phases~\cite{Senthil2003}. 
The FL$^*$ can be realized even for the gapless Kitaev spin liquid, 
because the density of states of the Majorana fermions is vanishing at low energy.
In the Kitaev-Kondo lattice, the topological $p$-wave superconductivity can be realized in an intermediate region of the 
Kondo coupling~\cite{Seifert2018,choi2018topological}.
Besides, superconductivity can be induced by non-spin-liquid order such as the valence bond solid in the
Shastry-Sutherland-Kondo lattice model~\cite{coleman2010frustration,su2015dimer}.
These theoretical studies would be relevant not only for fundamental understanding of frustrated 
Kondo systems but also for experiments of real materials.
For example, the $f$-electron comounds Yb$_2$Pt$_2$Pb and Ce$_2$Pt$_2$Pb 
are Shastry-Sutherland-Kondo systems where frustration
plays important roles~\cite{kim2008yb, kim2013spin, kim2011heavy}.
Furthermore, the Kitaev-Kondo model may be realized in a heterostructure of the Kitaev magnet 
$\alpha$-RuCl$_3$ and graphene,
where non-trivial interplay between the charge degrees of freedom and the frustration is expected
~\cite{Biswas2019-lc,Rossi2023-cu}.

In this work,
motivated by the theoretical and experimental developments of frustrated Kondo lattices, 
we investigate competition and cooperation of the Kondo effect and 
the spin nematic order which is a frustration induced spin order.
For a spin nematic state, a conventional dipole moment vanishes, $\expval{\bm{S}_i}=0$,
but a quadrupole moment has a non-zero expectation value.
In a spin-1/2 system, 
the spin nematic order is characterized by an inter-site spin quadrupole moment $(\mu, \nu = x,y,z)$ on a bond,
\begin{equation*}
  \expval{Q_{ij}^{\mu \nu}}
  \equiv \expval{S_i^{\mu} S_j^{\nu}} + \expval{S_i^{\nu} S_j^{\mu}} - \frac{2}{3}\expval{\bm{S}_i \cdot
  \bm{S}_j} \delta_{\mu\nu}.
\end{equation*}
Experimentally, there are several candidate materials for the spin nematic state such as the quasi
one-dimensional compound LiCuVO$_4$~\cite{Enderle2005,Orlova2017}.
The quasi two-dimensional systems $AA'$VO(PO$_4$)$_2$ ($AA'=$Pb$_2$, BaCd, etc.) 
are located close to the spin nematic state~\cite{Kaul2004,Nath2008}.
The spin-nematic order has been theoretically discussed in various models,
such as zigzag spin chain model~\cite{Chubukov1991-co,Hikihara2008-vm,Zhitomirsky2010-th,Nakamura2024},
triangular lattice multiple-spin-exchange model~\cite{Momoi2005-rj,Momoi2006-nc,Momoi2012-cb},
$J_1$-$J_2$ model~\cite{Shannon2006-oj,Shindou2009-fj,Jiang2023-ls},
and $J_1$-$J_2$-$J_3$ model~\cite{Sindzingre2010-ex,Iqbal2016-ns}.
In particular, in the $J_1$-$J_2$ model on the square lattice
with ferromagnetic $J_1$ and antiferromagnetic $J_2$ interactions, 
it was argued that the spin nematic state
is realized for a large range of magnetic fields including zero magnetic field~\cite{Shannon2006-oj}.
The spin nematic state in absence of a magnetic field can be well described by 
an SU(2) slave-boson theory with fractionalized spinon excitations~\cite{Shindou2009-fj}.
Within this theory, the spin nematic state is described as a resonating valence bond state, where
magnetic dipole moments are naturally zero and spin quadrupole moments can be
induced by the spinon triplet pairing and the ``spin-orbit" hopping. 
Although the zero field spin nematic state
in the $J_1$-$J_2$ model was questioned by the detailed numerical calculation~\cite{Jiang2023-ls}, 
the subtle energy competition and the previous theoretical results
imply that the system is located close to a spin nematic state~\cite{Shannon2006-oj,Shindou2009-fj},
where the spin nematicity might be enhanced by an additional interaction~\cite{Sato2013}.
Experimentally, a spin nematic state essentially at zero magnetic field was found in 
the square lattice iridate~\cite{Kim2024}.
These theoretical and experimental observations 
would imply that the $J_1$-$J_2$ model and its possible generalizations are 
promissing candiates for realization of the spin nematic state
at zero magnetic field.

Moreover, a junction of the spin nematic compound LiCuVO$_4$ and the metallic Pt has been fabricated~\cite{Hirobe2019}.
Spin transport through the interface between LiCuVO$_4$ and Pt
was experimentally investigated and 
it was theoertically analyzed within perturbative treatment of the Kondo coupling at the interface.
However,
a magnet and a metal can be tightly coupled in general as in the previously mentioned $\alpha$-RuCl$_3$/graphene
system~\cite{Biswas2019-lc,Rossi2023-cu}.
Therefore, it is interesting to expolre impacts of strong Kondo coupling and study 
basic relations of the spin nematic order and the Kondo effect.
%as a first step,
In this work, 
we introduce the $J_1$-$J_2$ Kondo lattice model on the square lattice as a minimal model 
and analyze it by using the 
slave boson mean field approximation.
We discuss stability of an FL$^*$ state and an HFL state, and emergence of superconductivity in between these
two phases.

This paper is organized as follows.
In Sec.~\ref{sec:level2}, we introduce the $J_1$-$J_2$ Kondo lattice model on the square lattice
and extend the previous slave boson theory for the spin model to the Kondo lattice.
In Sec.~\ref{sec:level3}, numerical results of the slave boson approximation are discussed for
a low filling case and a high filling case, respectively.
Finally, a summary is given in Sec.~\ref{sec:level4}.

%%%%%%%%%%%%%%%%%%%%%%%%%%%%%%%%%%%%%%%%%%%%%%%%%%%%%%%%%%%%%%%%%%%
\section{\label{sec:level2}MODEL AND slave-boson mean field approximation}
\begin{figure}[htb]
  \centering
  \includegraphics[width=7cm]{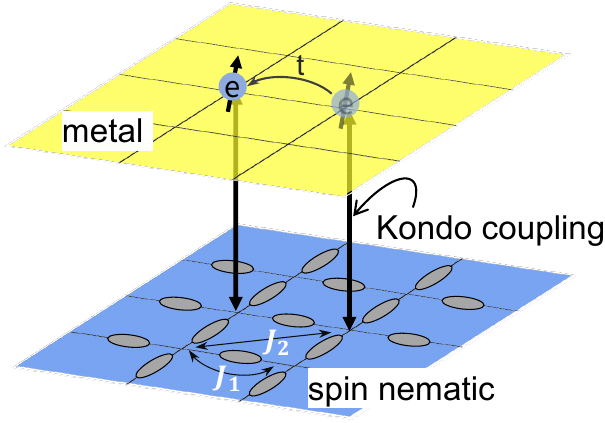}% Here is how to import EPS art
  \caption{The schematic picture of $J_1$-$J_2$ Kondo lattice model on the square lattice.
  The localized spins with the FM nearest neighbor interaction ($J_1$) and the AFM next nearest neighbor interaction ($J_2$) are 
  Kondo coupled to the conduction electrons with the hopping $t$.
  The ellipse on a bond represents the nematic order parameter.}
  \label{fig:1}
\end{figure}

We consider 
the $J_1$-$J_2$ Kondo lattice model (Fig.~\ref{fig:1}) on the two-dimensional square lattice,
\begin{equation}
  H = H_{t} + H_{K} + H_{J_1\mbox{-}J_2}.
\end{equation}
The first and second terms correspond to the standard Kondo lattice,
\begin{align}
  H_{t} &= -t \sum_{\expval{ij}} \qty(\cdia \cja + \mathrm{h.c.}),\\
  H_{K} &= J_{K} \sum_{i} \bm{s}_{i} \cdot \bm{S}_{i},
\end{align}
where $\cdia (\cia)$ creates (annihilates) a conduction electron at site $i$ with
spin $\alpha=\uparrow,\downarrow$.
The repeated spin indices are summed over in our notation.
$t$ is the hopping integral between the nearest neighbor sites and $J_K\geq0$ is the antiferromagnetic (AFM) Kondo 
coupling between the conduction electron spin 
$\bm{s}_i=c^{\dagger}_{i\alpha}(\bm{\sigma}_{\alpha\alpha'}/2)c_{i\alpha'}$ ($\bm{\sigma}$ are the Pauli matrices) 
and the $S=1/2$ localized spin $\bm{S}_{i}$.
The third term describes interactions between the localized spins, 
\begin{equation}
  H_{J_1\mbox{-}J_2} = -J_{1} \sum_{\expval{ij}} \bm{S}_{i} \cdot \bm{S}_{j}
                    + J_{2}\sum_{\expval{\expval{ij}}} \bm{S}_{i} \cdot \bm{S}_{j},
\end{equation}
where $\expval{ij}$ and $\expval{\expval{ij}}$ represent a pair of nearest neighbor sites
and second nearest neighbor sites, respectively. 
This spin model is called the $J_1$-$J_2$ model and has been extensively studied in the context of frustrated 
spin systems~\cite{Shannon2006-oj,Shindou2009-fj,Jiang2023-ls}. 
In this study, we are interested in a parameter region favoring the spin nematic state,
where the nearest neighbor 
interaction is ferromagnetic (FM) $-J_1\leq0$
and the next nearest neighbor interaction is antiferromagnetic $J_2\geq0$.
A uniform FM state is realized for $J_1\gg J_2$, while a stripe AFM state is stable for $J_1\ll J_2$.
In an intermediate region, the spin nematic state emerges as a result of the magnetic frustration~\cite{Shannon2006-oj,Shindou2009-fj,Jiang2023-ls}.
The interaction is parameterized as $J_1=\sin\theta, J_2=\cos\theta$,
and a wide range of the Kondo coupling $0\leq J_K\leq 7$ is considered,
where the energy unit is $t=1$.
The magnetic interactions $J_1,J_2$ and the electron hopping $t$ are supposed to be 
of the same order as in the previous studies for frustrated Kondo lattice models
~\cite{Senthil2003,coleman2010frustration,Vojta2018,Seifert2018,choi2018topological,Bernhard2011,Pixley2014,su2015dimer}.
If we use smaller (larger) values of $J_1, J_2$, the spin nematic state would become more unstable (stable) 
against $J_K$.
We discuss the low electron filling $n_{\mathrm{c}}=0.3$ and the high electron filling $n_{\mathrm{c}}=0.8$, respectively.

As was mentioned in the previous section,
the $J_1$-$J_2$ model is a prototypical model for spin nematic states 
and the stability of the spin nematic state could be approximately described by a slave-boson
mean field approach.
In the previous study based on the slave-boson mean field approximation, 
it was shown that the $J_1$-$J_2$ model at zero magnetic field exhibits
spin nematic states for $1.02\lesssim J_2/J_1\lesssim 1.29$~\cite{Shindou2009-fj}.
Within this approach,
the spin nematic state is characterized as a resonating valence bond state with a vanishing magnetic dipole moment 
and a non-zero quadrupole moment which arises from spinon hopping and pairing. 
The spinon band structures are classified by the pairing structures, and there are two stable pairing states. 
One is the helical (or Balian-Werthamer) spin nematic state which is analogous to the $^3$He-B phase 
with time-reversal symmetry~\cite{Leggett1975,Leggett2006}.
The other is the chiral spin nematic state (or Anderson-Brinkman-Morel) corresponding to the $^3$He-A phase where 
the time-reversal symmetry is broken.
These spin nematic states are stable for $\theta_{c1}=0.66\leq \theta\leq \theta_{c2}=0.76
(1.05\lesssim J_2/J_1\lesssim 1.29)$ and 
$\theta_{c2}=0.76\leq \theta\leq \theta_{c3}=0.775 (1.02\lesssim J_2/J_1\lesssim 1.05)$ within the slave-boson mean field approximation, respectively~\cite{Shindou2009-fj}.
Although the helical state is consistent with the exact diagonalization study~\cite{Shannon2006-oj}, 
it was pointed out that 
stability of the chiral state beyond the slave-boson mean field solution and associated small gauge fluctuations
is not clear~\cite{Shindou2009-fj}.
Here, we focus only on the parameter region $\theta_1\leq \theta\leq \theta_2$
where the helical spin nematic state is stable when $J_K=0$.
Note that time-reversal symmetry breaking may occur in a generalized model with additional terms such as the
multi-spin interactions~\cite{Shannon2006-oj,Sato2013}.

In this study, we use the slave-boson mean field approximation to describe both the spin nematic order 
and the Kondo effect on an equal footing.
We extend the previous slave boson theory for the spin model~\cite{Shindou2009-fj}
to the Kondo lattice model.
To this end, we decompose localized spin operators into spin-1/2 charge-neutral
spinons and impose a constraint on the spinon number at each site to exclude double
occupancy of spinons,
\begin{align}
  &\bm{S}_i = \frac{1}{2} f_{i \alpha}^{\dag} (\bm{\sigma})_{\alpha \beta} f_{i \beta}, \\
  &f_{i \alpha}^{\dag}  f_{i \alpha} = 1, ~ ~ f_{i \uparrow}^{\dag} f_{i \downarrow}^{\dag}=
  f_{i \downarrow} f_{i \uparrow} = 0
\end{align}
where $f_{i \alpha}^{\dag} (f_{i \alpha})$ is creation (annihilation) operators of a spinon
at site $i$.
These constraints are incorporated by the use of Lagrange multiplier $a_{i, \mu}$.

The spin interactions $J_K,J_1,J_2$ possess distinct properties (FM or AFM,
on-site or inter-site), and thus it is natural to perform different decompositions for each of
these interactions.
First, we decompose AFM on-site interaction $J_{K}$ as follows,
\begin{equation}
  \bm{s}_{i} \cdot \bm{S}_{i} = - \frac{1}{2} \qty(\cdia \fia) \qty(\fdib \cib)
  + \frac{1}{4} \cdia \cia
\label{eq:decK}
\end{equation}
where the constraint $f^{\dagger}_{i\alpha}f_{i\alpha}=1$ has been used.
The mean-field order parameter is the (dimensionless) Kondo hybridization
\begin{align}
V_i \equiv \expval{\fdia \cia}.
\end{align}
Next, we decompose the inter-site interactions $J_{1},J_2$.
We include pairing terms in the decomposing procedure by 
following the previous study~\cite{Shindou2009-fj}.
The FM interaction $J_1$ can be written as
\begin{align}
  - 4 \bm{S}_{i} \cdot \bm{S}_{j} &= - \sum_{\mu=x,y,z}\qty(\fdia \qty(\sigma_{\mu})_{\alpha \beta} 
  \fjb) \qty(\fdjg \qty(\sigma_{\mu})_{\gamma \eta} \fie) \nonumber \\
  & - \sum_{\mu=x,y,z}\qty(\fdia \qty(\sigma_y \sigma_{\mu})_{\alpha \beta} \fdjb) 
  \qty(\fjg \qty(\sigma_{\mu} \sigma_y)_{\gamma \eta} \fie),
\label{eq:dec1}
\end{align}
while the AFM interaction $J_2$ is
\begin{align}
  4 \bm{S}_{i} \cdot \bm{S}_{j} =& - \qty(\fdia \fja) \qty(\fdjb \fia) \nonumber \\
  & - \qty(\fdia (\sigma_y)_{\alpha \beta} \fdjb) \qty(\fjg (\sigma_y)_{\gamma \eta} \fie).
\label{eq:dec2}
\end{align}
Correspondingly,
the mean-field order parameters for the nearest neighbor bond with the FM interaction $J_1$
are spinon ``spin-orbit" hopping and spin-triplet pairing,
\begin{align}
E_{ij,\mu} \equiv\expval{\fdia (\sigma_{\mu})_{\alpha \beta} \fjb}, \quad
D_{ij,\mu} \equiv 
\expval{\fia (i \sigma_{\mu} \sigma_y)_{\alpha \beta} \fjb}.
\end{align}
For the next nearest neighbor bond with the AFM interaction $J_2$, the mean-field
parameters are spinon ``spin-conserving" hopping and spin-singlet pairing,
\begin{align}
\chi_{ij} \equiv \expval{\fdia \fja},\quad
\eta_{ij} \equiv \expval{\fia (i \sigma_y)_{\alpha \beta} \fjb}.
\end{align}
Note that the spinon pairing order parameters $(\eta,\bm{D})$ are analogous to the $d$-vector of a superconductor,
$(d_0,\bm{d})$, where a gap function is written as 
$\Delta_{\alpha\beta}=d_{\mu}(i\sigma_{\mu}\sigma_y)_{\alpha\beta}$~\cite{Leggett1975,Leggett2006}.
The mean-field order parameters are directly related to the spin nematic order as
\begin{align}
Q_{ij,\mu\nu}&=-\frac{1}{2}\left( E_{ij,\mu}E_{ij,\nu}^* -\frac{1}{3}\delta_{\mu\nu}|E_{ij,\mu}|^2 +\mbox{(h.c.)}\right)
\nonumber\\
&\quad -\frac{1}{2}\left( D_{ij,\mu}D_{ij,\nu}^* -\frac{1}{3}\delta_{\mu\nu}|D_{ij,\mu}|^2 +\mbox{(h.c.)}\right).
\end{align}
From this expression, one can immediately see that a spin nematic state within the slave-boson mean field
approximation is described by spinons with the spin-orbit hopping or the spin-triplet pairing.
The helical state and chiral state have non-zero $D_{ij,\mu}$ with different symmetries
as will be explained later.

Imposing the constraint on the conduction electron filling,
\begin{equation}
  \expval{\cdia \cia} - n_{\mathrm{c}} = 0,
\end{equation}
and using the decomposition mentioned above, the total mean-field Hamiltonian is written as
\begin{align}
  \label{eq:H_MF(real space)}
  H_{\mathrm{MF}} &= -\frac{J_1}{4} \sum_{\expval{ij}} \Bigl[E_{ij, \mu}^{\ast} \fdia 
  (\sigma_{\mu})_{\alpha \beta}\fjb \nonumber \\
  &\quad \qquad \qquad+ D_{ji, \mu} \fdia (-i \sigma_y \sigma_{\mu})_{\alpha \beta} \fdjb + \mathrm{h.c.} \Bigr] \nonumber \\
  &-\frac{J_2}{4} \sum_{\expval{\expval{ij}}} \Bigl[\chi_{ji} \fdia \fja + \eta_{ij} \fdia (-i \sigma_y)_{\alpha \beta}
  \fdjb + \mathrm{h.c.}\Bigr] \nonumber \\
  &- \frac{J_{K}}{2} \sum_{i} \Bigl( V_i \cdia \fia + \mathrm{h.c.}\Bigr)
  - t \sum_{\expval{ij}} \qty(\cdia \cja + \mathrm{h.c.}) \nonumber \\
  & - \mu  \sum_i \Bigl( \cdia \cia - n_{\mathrm{c}} \Bigr) + \sum_{i} a_{i,z}\qty(\fdia \fia - 1) \nonumber \\
  & + \sum_{i} \qty(a_{i,x} f_{i \uparrow}^{\dag} f_{i \downarrow}^{\dag} 
  + a_{i,y} f_{i \downarrow} f_{i \uparrow}) + E_0 \\
  E_0 =& \frac{J_1}{4} \sum_{\expval{ij}} \qty(|\bm{E}_{ij}|^2 + |\bm{D}_{ij}|^2)
  + \frac{J_2}{4} \sum_{\expval{\expval{ij}}} \qty(|\chi_{ij}|^2 + |\eta_{ij}|^2) \nonumber \\
  & + \frac{J_{K}}{2} \sum_i |V_i|^2.
\end{align}

In this study,
we assume that 
the translation symmetry is preserved in the ground state. 
Then, the order parameters depend only on the relative positions of two sites,
namely, $E_{ij,\mu}
(D_{ij,\mu}) = E_{\hat{x},\mu} (D_{\hat{x},\mu})$ ($x$ bond) or $E_{\hat{y},\mu} (D_{\hat{y},\mu})$
($y$ bond), $\chi_{ij} (\eta_{ij}) = \chi_+ (\eta_+)$ ($x+y$ bond) or $\chi_- (\eta_-)$ ($x-y$ bond),
$V_i = V, a_{i, \mu} = a_{\mu}$.
By applying a Fourier transformation to the mean-field Hamiltonian (\ref{eq:H_MF(real space)}) and
using the Nambu spinor $\Psi_{\bm{k}}=(f_{k\uparrow},f_{k\downarrow},c_{k\uparrow},c_{k\downarrow},
f_{-k\uparrow}^{\dagger},f_{-k\downarrow}^{\dagger},
c_{-k\uparrow}^{\dagger},c_{-k\downarrow}^{\dagger})$, the Hamiltonian can be written as
\begin{align}
  H_{\mathrm{MF}} &= \sum_{\bm{k}} \Psi_{\bm{k}}^{\dag} \hat{H}_{\mathrm{MF}} \Psi_{\bm{k}} + \tilde{E_0}, \\
  \tilde{E_0}/2N &= \frac{J_1}{4} \sum_{\delta=\hat{x},\hat{y}} \qty(|\bm{E}_{\delta}|^2  
  + |\bm{D}_{\delta}|^2) \nonumber \\
  + \frac{J_2}{4} &\sum_{\xi=+,-} \qty(|\chi_{\xi}|^2 + |\eta_{\xi}|^2) +
  \frac{J_K |V|^2}{4} +\frac{\mu}{2} \qty(n_{\mathrm{c}}-1).
\end{align}
where $\hat{H}_{\mathrm{MF}}$ is the corresponding $8\times 8$ Bogoliubov-de Gennes Hamiltonian 
and $N$ is the number of sites.

We explore mean-field solutions of the fields $(E,D,\chi,\eta,V,a,\mu)$ that minimize the ground state energy per site
\begin{equation}
  E_{\mathrm{MF}} = \frac{1}{2N} \sum_{\bm{k}} \sum_{n=1}^{4} \qty(E_{n}(\bm{k}) + \tilde{E_0}), 
\label{eq:E0}
\end{equation}
where $E_{n}(\bm{k})$ are Bogoliubov quasiparticle dispersions obtained by numerical diagonalization
of the matrix $\hat{H}_{\mathrm{MF}}$ and they are
arranged in ascending order.
If there is a solution with both $D\neq0$ (or $E\neq0$) and $V\neq0$, the system will exhibit the spin nematic order and
the Kondo effect simultaneously.
In such a state, the localized spins and the conduction electrons are tighly coupled and 
Cooper pairing of the conduction electrons can be induced by the spinon pairing through the non-zero Kondo 
hybridization,
which is a non-trivial interplay between the spin nematic order and the Kondo effect. 
This can be examined by 
directly calculating the Cooper pairing amplitude with use of the slave boson mean field solutions.
Once we obtain a mean-field solution, the pairing amplitudes of the conduction electrons are evaluated as,
\begin{align}
  \Delta_{\delta, \mu}^{\mathrm{tri}} &= \expval{\cia (i \sigma_{\mu} \sigma_y)_{\alpha \beta} c_{i+\delta \beta}}, \\
  \Delta_{\pm}^{\mathrm{sin}} &= \expval{\cia (i \sigma_y)_{\alpha \beta} c_{i+\hat{x}\pm \hat{y} \beta}}
\end{align}
for $\delta = \hat{x}, \hat{y}$.
The pairing amplitude $\Delta^{\mathrm{tri}}$ describes triplet pairing on horizontal or vertical bonds corresponding
to the FM $J_1$ spin interaction
and $\Delta^{\mathrm{sin}}$ does singlet pairing on diagonal bonds corresponding to the AFM $J_2$ spin interaction.
Depending on the spinon pairing structures, different kinds of electron pairings can be realized;
(i) purely spin-singlet $d$-wave pairing $\Delta^{\mathrm{sin}}\neq0, \Delta^{\mathrm{tri}}=0$,
(ii) purely  spin-triplet $p$-wave pairing $\Delta^{\mathrm{sin}}=0, \Delta^{\mathrm{tri}}\neq0$,
and (iii) $p+d$-wave mixed pairing $\Delta^{\mathrm{sin}}\neq0, \Delta^{\mathrm{tri}}\neq0$.
We will discuss possible stabilization of these states in the next section.

%%%%%%%%%%%%%%%%%%%%%%%%%%%%%%%%%%%%%%%%%%%%%%%%%%%%%%%%%%%%%%%%%%
\section{\label{sec:level3}numerical RESULTS}

We investigate competition and cooperation of the spin-nematic order and the Kondo effect
at zero temperature by calculating the mean field energy Eq.~\eqref{eq:E0}. 
In this study, we focus on 
the four stable states found in the previous study of
the $J_1$-$J_2$ spin model without the conduction electrons~\cite{Shindou2009-fj}.
These states were obtained by the extensive parameter optimization and hence are the most relevant candidates
for the ground state in the present $J_1$-$J_2$ Kondo lattice model.
The first state for the spin model (at $J_K=0$) is
the helical spin nematic state with time-reversal symmetry which has the order parameters~\cite{Shindou2009-fj}
\begin{align}
  &D_{\hat{x}, x} = D_{\hat{y},y} = D ~(\mathrm{real}), \quad \chi_+ = \chi_- = \chi ~(\mathrm{real}), \nonumber \\
  &\eta_+ = - \eta_- = \eta ~(\mathrm{real}), ~ \mathrm{others} = 0.
\label{eq:helical_ansatz}
\end{align}
The second state is the chiral spin nematic state with broken time reversal symmetry,
\begin{align}
  &D_{\hat{x}, z} = -iD_{\hat{y},z} = D ~(\mathrm{real}), \quad \chi_+ = \chi_- = \chi ~(\mathrm{real}), \nonumber \\
  &\mathrm{others} = 0.
\end{align}
The third one is
the collinear spin nematic state,
\begin{align}
  &D_{\hat{x}, z} = D_{\hat{y},z} = D ~(\mathrm{real}), \quad \chi_+ = \chi_- = \chi ~(\mathrm{real}), \nonumber \\
  &\eta_+ = - \eta_- = \eta ~(\mathrm{real}), ~ a_x \neq 0, ~ \mathrm{others} = 0,
\end{align}
and the fourth is the $d$-wave state (including the $\pi$-flux state with $\chi= \eta$),
\begin{align}
  &\chi_+ = \chi_- = \chi ~(\mathrm{real}), \quad \eta_+ = - \eta_- = \eta ~(\mathrm{real}), \nonumber \\
  &\mathrm{others} = 0.
\end{align}
These ansatz states are exteded to the Kondo lattice model where $V$ and $\mu$ are also simultaneously
calculated.
As another stable solution in the $J_1$-$J_2$ spin model, there is a flat-band state which is expected
to be a ferromagnetic state when projected onto the physical Hilbert space.
However, it turns out that this solution does not converge once the Kondo coupling $J_K$ is introduced.
Besides, according to the previous study, the ferromagnetic state in the standard Kondo lattice 
(without $J_1$ and $J_2$) at a low filling region is not stable and the paramagnetic HFL
state is realized for moderate Kondo coupling $J_K$~\cite{bernhard2015coexistence}. 
This would imply that the ferromagnetic state
is not stable  also in the present $J_1$-$J_2$ Kondo lattice for moderate $J_K$.
In this study, we are primarily interested in such a parameter region and hence
restrict ourselves to the above four stable states as candidates of the ground state.
In the following, we present the numerical results of 
the variational parameters $(E,D,\chi,\eta,V,a,\mu)$ 
obtained by minimizing the ground state energy Eq.~\eqref{eq:E0}
and summarize them into a ground state phase diagram.

%%%%%%%%%%%%%%%%%%%%%%%%%%%%%%%%
\subsection{Low conduction electron filling ($n_c=0.3$)}

\begin{figure}[htb]
  \centering
  \begin{subfigure}[b]{0.35\textwidth}
      \centering
      \includegraphics[width=\textwidth]{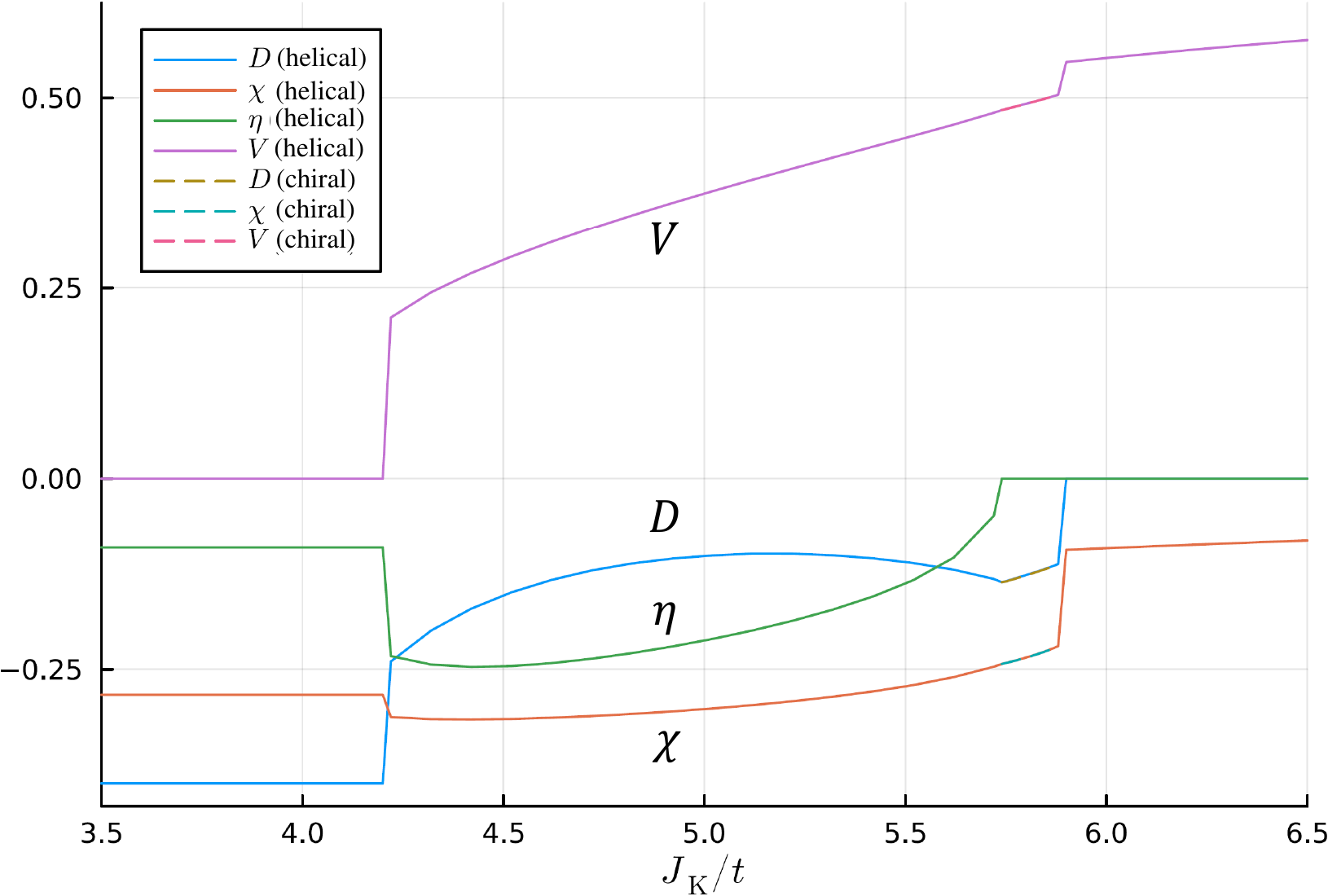}
      \caption{}
      \label{fig:2-a}
  \end{subfigure}
  \hfill
  \begin{subfigure}[b]{0.35\textwidth}
      \centering
      \includegraphics[width=\textwidth]{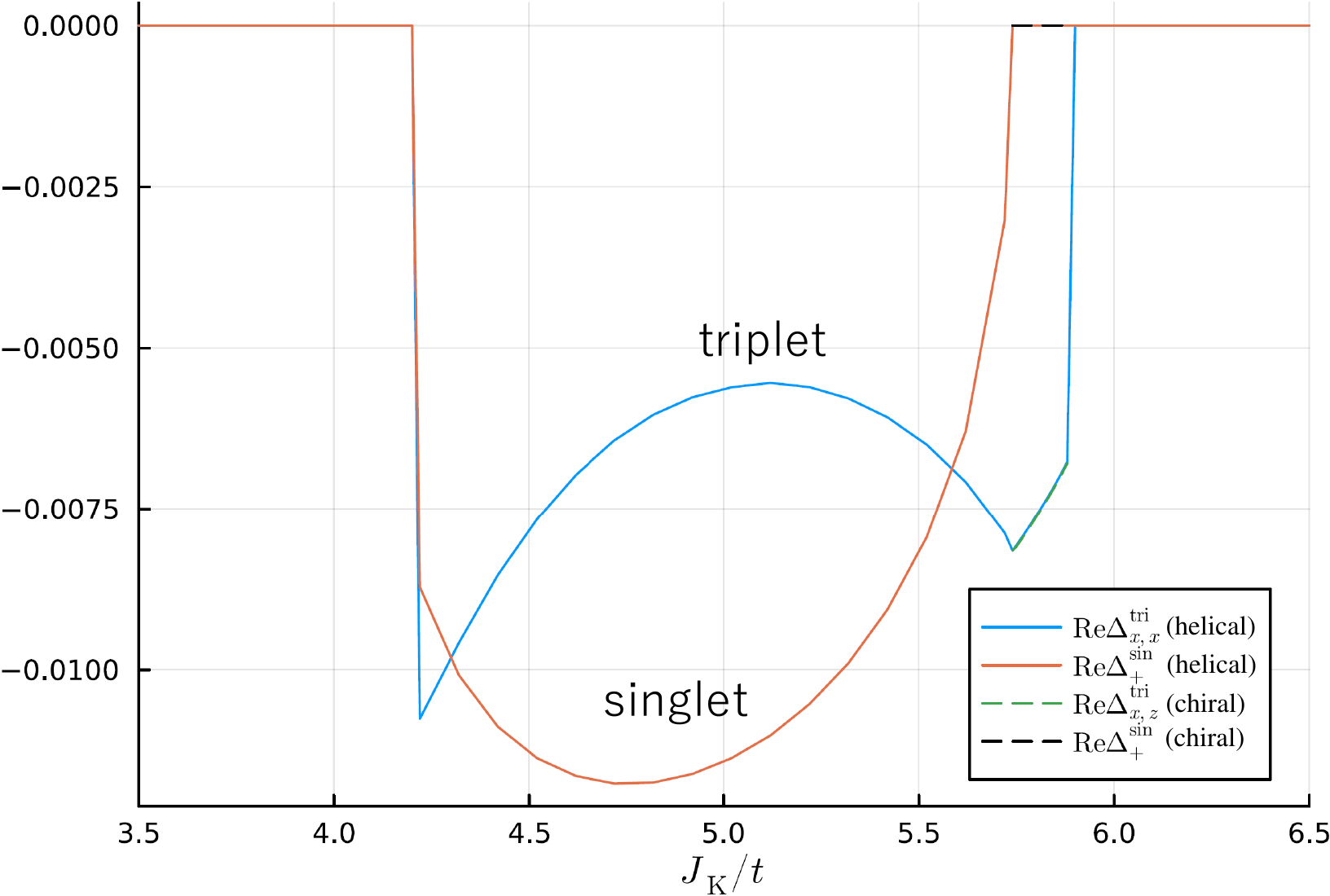}
      \caption{}
      \label{fig:2-b}
  \end{subfigure}
  \caption{(a) Mean field order parameters and (b) superconducting pairing amplitudes with
  $\theta=0.76,~ n_{\mathrm{c}} = 0.3$. 
  Note that the helical superconducting state and the chiral superconducting state are degenerate
  and the order parameters of these two states coincide for $5.7\lesssim J_K\lesssim 5.8$.
  (The order parameters for the chiral state are shown by the dashed lines only for this range of $J_K$.)}
  \label{fig:2}
\end{figure}

First, we consider low conduction electron filling $n_c=0.3$ and discuss high filling case in the next section.
We show typical dependence of the order parameters on the Kondo coupling at $\theta=0.76,~ n_{\mathrm{c}} = 0.3$
in Fig.~\ref{fig:2} (a) (The spin interactions are parametrized as $J_1 = \sin \theta$,
$J_2 = \cos \theta$ with the energy unit $t=1$.)
There is no coupling between the localized spins and the conduction electrons when $J_K=0$,
and the localized spins exhibit the spin nematicity.
In a standard Kondo lattice (without $J_1,J_2$), they will be tightly coupled to exhibit RKKY-mediated magnetism 
once a small $J_K$ is introduced.
In sharp contrast, in the present frustrated system,  
the Kondo hybridization parameter remains zero, $V=0$, and 
the localized spins and the conduction electrons are effectively decoupled for a finite range of the Kondo coupling
up to a critical value $J_K\simeq 4.2$.
This phase can be regarded as an FL$^*$,
where the Fermi surface consists only of the conduction electrons without localized spins.
The localized spins are essentially decoupled from the conduction electrons and the spinons are deconfined.
The FL$^*$ is a non-Fermi liquid and clearly distinguished
from the standard Fermi liquid where the Fermi surface volume contains both the conduction electrons
and the localized spins~\cite{Senthil2003,coleman2010frustration,Vojta2018}.
The stability of the FL$^*$ is often supported by an energy gap of the localized spins such as
a $\mathbb{Z}_2$ spin liquid~\cite{Senthil2003}, 
but FL$^*$ can be stable even for a gapless Kitaev spin liquid with a vanishing density
of states of the gapless Majorana excitations at low energy~\cite{Seifert2018,choi2018topological}.
In the present system, the spinons are gapped for both the helical state and the chiral state similarly
to the fully gapped Bogoliubov quasiparticles in the $^3$He-B and $^3$He-A phases confined 
in two dimensions~\cite{Shindou2009-fj}.
Beyond the mean field approximation, there are possible gapless excitations for the spin nematic states,
namely, gauge fluctuations and order parameter fluctuations.
It was argued that the gauge fields are gapped for the present spin nematic states because of the SU(2) flux 
Higgs mechanism
or the U(1) Chern-Simons mechanism~\cite{Shindou2009-fj,Wen2007,Wen1991,Wen2002,Mudry1994}, 
and therefore the gauge fluctuations would not destroy
the FL$^*$ state found in the mean field approximation.
On the other hand, some of the order parameter fluctuations remain gapless 
in a multi-component pairing state even when the gauge fields are gapped.
%and they might affect the FL$^*$.
%However, we expect that their influence is small and the FL$^*$ is stable because their density of states
%is vanishing at low energy similary to the above mentioned FL$^*$ in the Kitaev-Kondo lattice with 
%the gapless excitations.
However, these gapless bosons do not destabilize the spin nematic order in the $J_1$-$J_2$ model,
and hence their influence on the stability of the FL$^*$ would also be irrelevant.

For a sufficiently large Kondo interaction, the spin nematic bond-order is suppressed and 
the on-site Kondo singlet is formed between the conduction electrons and the localized spins.
We note that existence of a HFL state depends on model parameters and it can be absent 
even for a large $J_K$  in a frustrated Kondo lattice
especially at low filling as discussed in the previous studies~\cite{su2015dimer,Seifert2018}.
In the present system, the conduction electron filling is relatively low $n_c=0.3$, for which 
the HFL state is naively expected to be less stable.
The existence of the HFL state in this system would imply strong competition 
between the $J_{1,2}$ and $J_K$ interactions.
The HFL remains to be stable at higher filling ($n_c=0.8$) as will be discussed later.
Beyond the mean field approximation, there can be U(1) gauge fluctuations associated 
with the spinon mean field.
In this context, the Kondo hybridization $V\sim f^{\dagger}c$ behaves as a bosonic Higgs field 
with the unit gauge charge corresponding to the spinon gauge charge (see also Appendix~\ref{app:su2}).
When the Higgs condensation takes place, 
the U(1) gauge field will be gapped and the spinon gauge charge will be confined
~\cite{Fradkin2013,FradkinShenker1979,Sachdev2023,Senthil2004}.
(Note that the Higgs phase and the confinement phase are smoothly connected in the compact U(1) Higgs model.)
The spinons alone do not carry a definite gauge charge in presence of the Higgs condensation $V$
and only the total charge of the spinon and conduction
electron is well defined corresponding to the large Fermi surface.

Despite the strong competition of the interactions, there can be cooperative phenomena.
In an intermediate $J_K$ region, a first-order phase transition occurs from the FL$^*$ state 
to a phase where the helical spin nematic order and the Kondo
hybridization coexist.
In this phase, the spinon pairing $\eta,D\sim\expval{ff}$ affects the conduction electrons through the Kondo hybridization $V\sim \expval{f^{\dagger}c}$,
leading to a non-zero electron pairing $\Delta\sim\expval{cc}$ as shown in Fig.~\ref{fig:2} (b).
At the present value of the spin interactions, $J_2/J_1\simeq 1.05$,
both of the spinon singlet order parameter $\eta$ and the triplet order parameter $D$ are non-zero, and 
correspondingly, spin-singlet pairing and triplet pairing of the conduction electrons can coexist at $J_K\neq0$.
The electron spin-singlet component $\Delta^{\mathrm{sin}}$ is induced by the spinon singlet-pairing order 
parameter $\eta$ on the diagonal bond corresponding to the AFM interaction $J_2$,
and the resulting superconducting symmetry is the $d_{xy}$-wave symmetry.
The electron triplet pairing component $\Delta^{\mathrm{tri}}$ arises from the spinon pairing $D$ corresponding
to the FM interaction $J_1$,
and the superconducting symmetry is a helical state whose $d$-vector
$(\Delta=\bm{d}\cdot i\bm{\sigma}\sigma_y)$ is given by $\bm{d}\sim (k_x,k_y,0)$ around the $\Gamma$-point.
This $p+d$-wave superconducting state preserves the time-reversal symmetry
and the Bogoliubov quasiparticles are fully gapped. 
We can consider stability of this superconducting state against the gauge fluctuations 
beyond the mean field approximation.
First,
the spinon mean field has a non-colinear SU(2) flux which makes the gauge field massive
as in the helical nematic state of the localized spins.
Besides, the gauge fluctuations are also quenched by the Higgs condensation $V\neq0$ (Appendix~\ref{app:su2}).
Therefore, there are no gapless gauge fluctuations and the $p+d$-wave superconductivity is stable.

As we further increase $J_K$ at fixed $J_2/J_1\simeq 1.05$, 
the singlet pairing component $\Delta^{\rm sin}$ continuously goes to zero, resulting in
purely triplet pairing superconductivity.
Although the helical nematic state with $\eta=0$ in the $J_1$-$J_2$ spin model is ${\mathbb Z}_2$ 
topologically trivial~\cite{Shindou2009-fj}, 
we find that the pure triplet helical superconducting state in the Kondo $J_1$-$J_2$ model is topologically non-trivial.
It is known that
the ${\mathbb Z}_2$ topological number for the helical superconducting state 
can be evaluated from the underlying Fermi surface structure 
in the non-superconducting state~\cite{Sato2017,Sato2009}.
When the nematic pairing mean field is turned off, ${\bm D}=0$,
the underlying large Fermi surface with $V\neq0$
surrounds only a single time-reversal invariant point in the Brillouin zone, 
in contrast to the underlying spinon Fermi suface with $V=0$ which surrounds two time-reversal invariant points.
This implies that the ${\mathbb Z}_2$ topological number in the helical superconducting state is odd.

We find that the ground state energy of the pure helical superconducting state is degenerate 
with that of another pure triplet
state, the chiral superconducting state, induced by the chiral spin nematic order.
Similar degeneracy for the helical and chiral spin nematic states were found in the $J_1$-$J_2$ spin model
without the conduction electrons
for $\theta_2\leq \theta \leq \theta_3 (1.02\leq J_2/J_1\leq 1.05)$,
where the spinon singlet pairing is vanishing ($\eta=0$) but the spinon spectrum remains gapped.
Beyond the mean field approximation, however, it was argued that the U(1) gauge fluctuations 
(especially the monopole fluctuations) lead to the spinon confinement when $\eta=0$
and the pure helical spin nematic state is unstable~\cite{Shindou2009-fj,Wen2007,
Wen1991,Wen2002,Mudry1994}.
However, similarly to the HFL state,
in the pure $(\eta=0)$ helical superconducting state of the present Kondo $J_1$-$J_2$ model,
the Kondo hybridization $V\sim f^{\dagger}c$ behaves as a Higgs field and hence 
the U(1) gauge field becomes massive 
~\cite{Fradkin2013,FradkinShenker1979,Sachdev2023,Senthil2004}.
This means that the pure helical state is stabilized by the Kondo hybridization.
On the other hand, the chiral spin nematic state with $\eta=0$ in the $J_1$-$J_2$ spin model is stable,
because the gauge fluctuations are massive due to the Chern-Simons mechanism~\cite{Shindou2009-fj}.
In the $J_1$-$J_2$ Kondo lattice,
the spinon gauge charge is no longer a good quantum number in presence of $V\neq0$
and correspondingly the quantized spinon Hall effect is absent, 
which implies that  there are no Chern-Simons terms at low energy.
However, the Kondo hybridization as a Higgs condensation induces masses for all the gauge fields in the
chiral state (see also Appendix~\ref{app:su2}).
Therefore, the chiral superconducting state is also stable to the gauge fluctuations.
To clarify which of the two superconducting states is more stable,
it is necessary to go beyond the mean field approximation and stability analysis of the gauge fluctuations.

Similar competition and cooperation of the spin nematic order and the Kondo effect can be seen for 
other values of the spin interactions $J_1,J_2$, but details depend on the ratio $J_2/J_1$.
When $1.13\lesssim J_2/J_1\lesssim 1.21$, the triplet component of the $p+d$-wave superconducting state 
becomes smaller rapidly as $J_K$ increases and the purely $d$-wave superconductivity is stabilized.
By further increasing $J_K$, there is a reentrant phase transition to the $p+d$-wave state and then
the system enters the HFL phase.
When $J_2/J_1\gtrsim 1.21$, the $p+d$-wave state is not stable for any $J_K$ 
and only the purely $d$-wave state is realized.
Note that, in the large $J_2/J_1$ region, the spin nematic order is suppressed ($E,D=0$)
and the corresponding spin state is a spin-singlet spinon-paired state only with
$\chi,\eta\neq0$ (generally $\chi\neq \eta$).
This state has
Dirac fermion excitations around nodes of the gap function and
remaining U(1) gauge degrees of freedom due to the colinear SU(2) gauge flux~\cite{Wen2007,
Wen1991,Wen2002,Mudry1994}.
Although the stability of such a state in the $J_1$-$J_2$ model is a subtle issue, 
the U(1) gauge fluctuations will be suppressed by the Kondo hybridization $V\sim\expval{f^{\dagger}c}\neq0$
as a Higgs field in the present Kondo $J_1$-$J_2$ model.
(Note that the U(1) gauge charge of 
the bosonic field $V$ is one corresponding to the spinon gauge charge, 
and the Higgs region and the confinement region are smoothly connected 
~\cite{Fradkin2013,FradkinShenker1979,Sachdev2023,Senthil2004}.)
Therefore, the $d$-wave superconducting state in our model is a stable phase within the slave boson approximation.

\begin{figure}[htb]
  \centering
  \includegraphics[width=8cm]{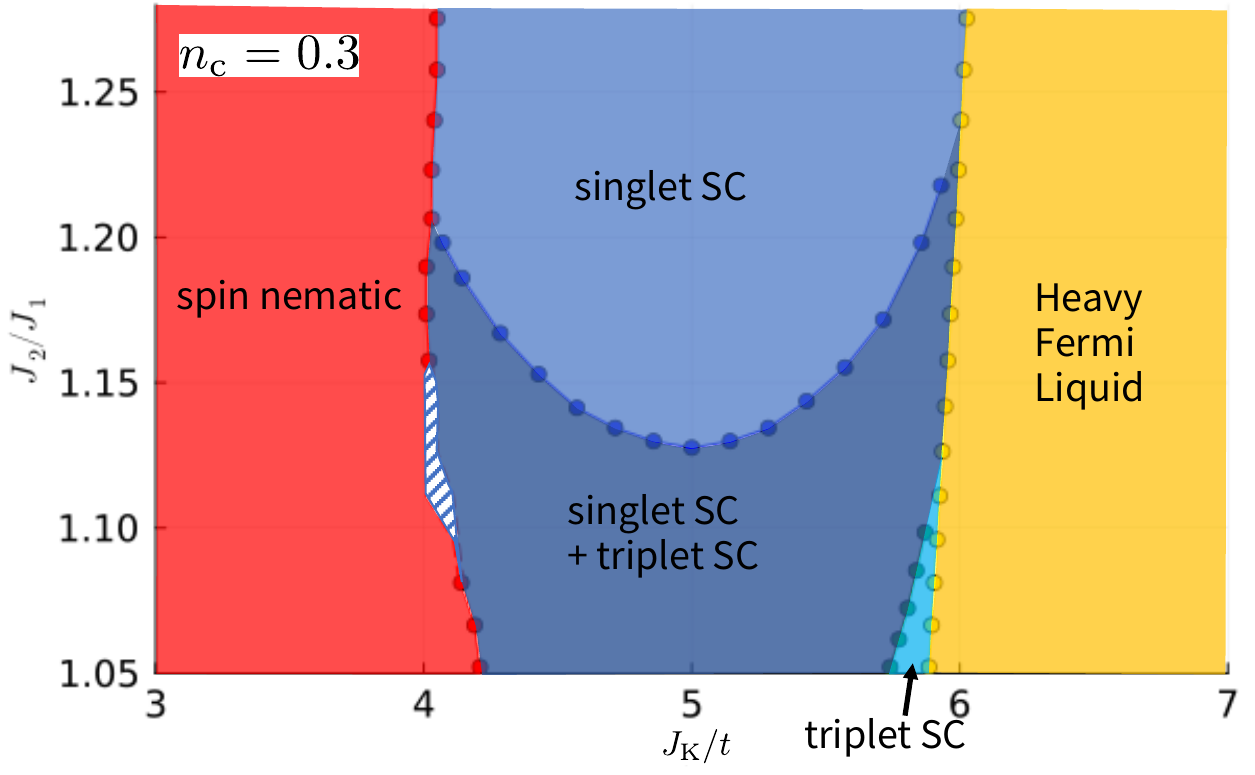}% Here is how to import EPS art
  \caption{The ground state phase diagram of the $J_1$-$J_2$ Kondo lattice model at $n_\mathrm{c}=0.3$.
For small $J_K$, the spin nematic state (red region) remains stable and it is regarded as the FL$^*$ state.  
For large $J_K$, the spin nematicity is destabilized and the paramagnetic heavy Fermi liquid (yellow region) becomes stable.
In the intermediate region of $J_K$, several superconducting states (SC) are found within the slave-boson mean field approximation.
Good convergence is not obtained for the shaded small region around $J_K=4$.} 
  \label{fig:3}
\end{figure}

These behaviors of the order parameters are summarized in the phase diagram (Fig.\ref{fig:3}).
As mentioned above,
in the range of $J_2/J_1$ where helical spin nematic order is realized at $J_K=0$, 
the spin-nematic state remains stable and the FL$^*$ state is stablized for small $J_K\neq0$.
As $J_K$ increases, 
the superconducting states are observed in the intermediate range, and the phase transition to the 
HFL state takes place at large $J_K$.

One interesting feature of this phase diagram is that there exists robust coexistence of the spin-singlet
pairing and the spin-triplet pairing of the conduction electrons, despite the total Hamiltonian
being fully SU(2) symmetric.
Although this is a direct consequence of the helical spin nematic state with both $\eta\neq0,\bm{D}\neq0$
arising from the spin interactions $J_1,J_2$,
it is non-trivial whether or not a singlet-triplet mixed superconducting state is stabilized when the localized spins
and the conduction electrons are coupled by $J_K$.
Indeed, the Bogoliubov quasiparticles are fully gapped by the triplet gap pairing $\Delta^{\mathrm{tri}}$
and one might naively expect that there is no room to have the singlet pairing $\Delta^{\mathrm{sin}}$.
This can be compared to the standard classification scheme of superconductivity,
where superconducting order parameters are classified by point group symmetry of the system 
considered~\cite{Sigrist1991}.
In this scheme, 
it is often assumed that one of the different pairing symmetries is realized,
and spin-singlet pairing and spin-tripet pairing do not coexist 
as long as there is no explicit spin SU(2) rotation symmetry breaking. 
Note that mere absence of spin SU(2) symmetry by e.g. an external magnetic field, a magnetic order,
and an asymmetric spin-orbit interaction
does not necessarily lead to coexistence of quantitatively comparable $\Delta^{\mathrm{sin}}$ and 
$\Delta^{\mathrm{tri}}$~\cite{Bauer2012}, 
and existence of pairing interactions for both channels is crucial~\cite{Tada2009}.
In principle, coexistence of different pairing symmetries is possible below a transition temperature
and a careful examiniation of the free energy (or the ground state energy) is necessary.
The coexistence in the present model is a non-trivial consequence of the frustration of the interactions
$J_1, J_2,J_K$.

\begin{figure}[htb]
  \centering
  \includegraphics[width=6cm]{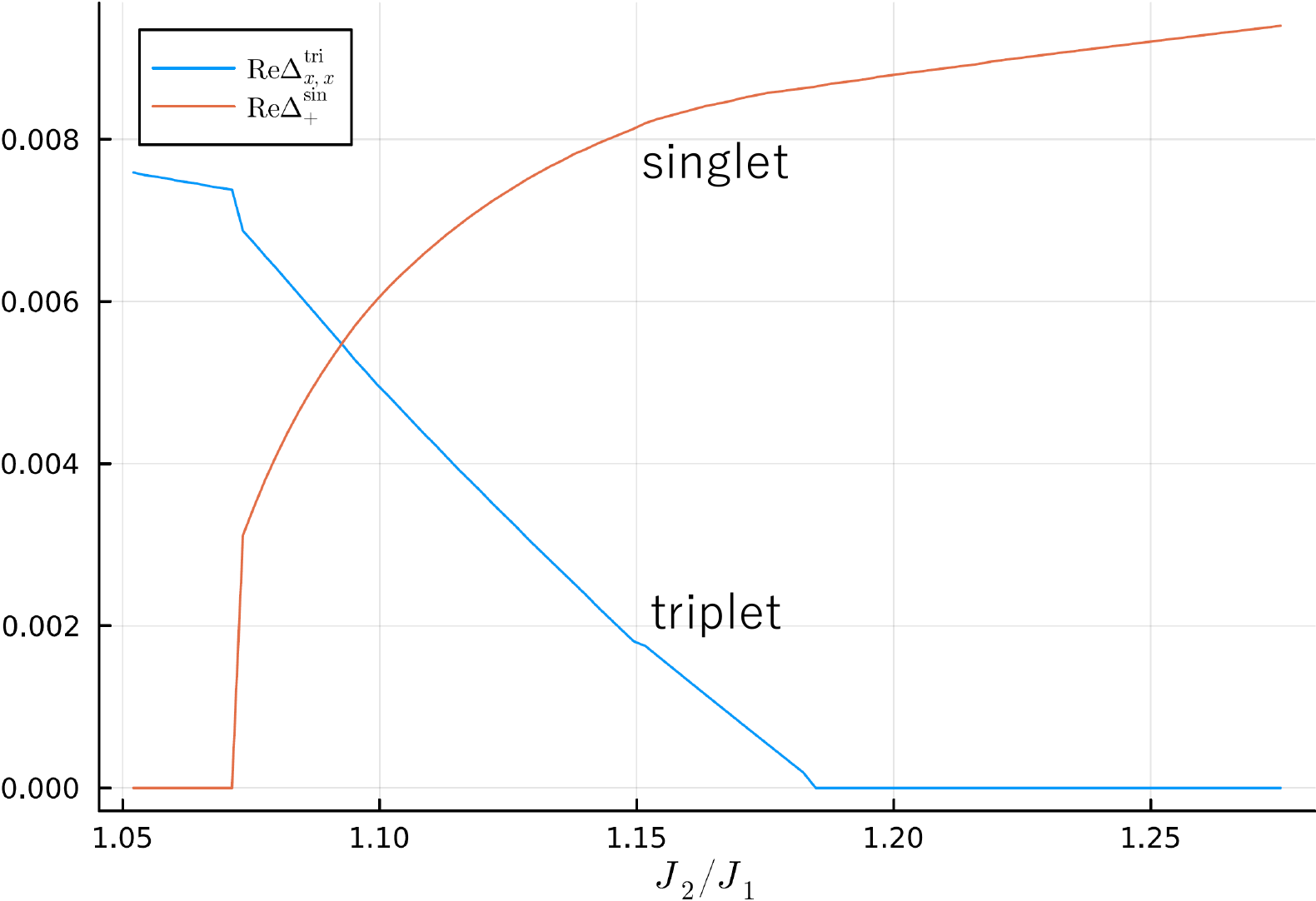}% Here is how to import EPS art
  \caption{The $J_2/J_1$ dependence of the superconducting pairing amplitudes at $J_K = 5.8, n_{\mathrm{c}} = 0.3$.
  The triplet pairing (blue) shows a linear growth around $(J_2/J_1)_{c2}\simeq 1.18$, 
  whereas the singlet pairing (orange) exhibits a square root like behavior around $(J_2/J_1)_{c1}\simeq 1.02$.
Note that it is difficult to fully identify the exact behaviors around the transition points due to 
discreteness of the data points and slow numerical convergence.}
  \label{fig:4}
\end{figure}

\begin{figure}[htbt]
  \centering
  \begin{subfigure}[b]{0.23\textwidth}
      \centering
      \includegraphics[width=\textwidth]{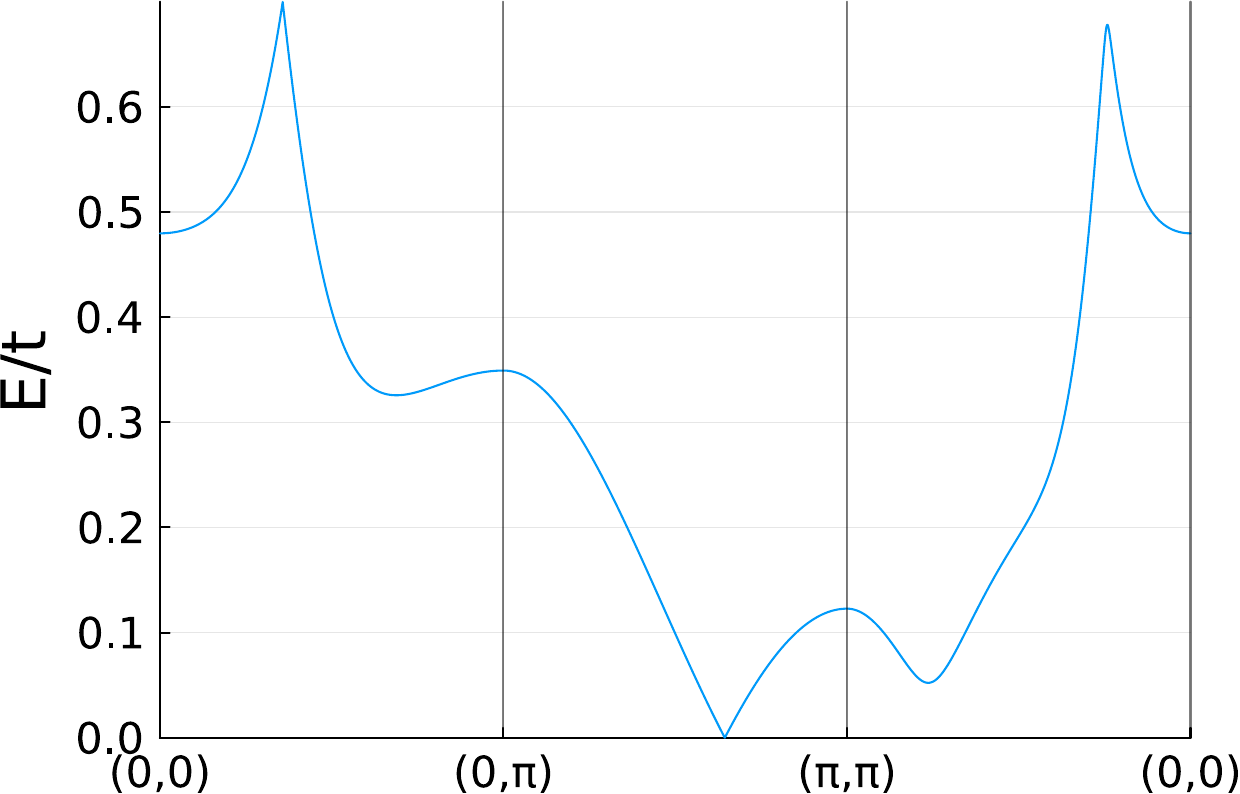}
      \caption{}
      \label{fig:5-a}
  \end{subfigure}
  % \hfill
  \begin{subfigure}[b]{0.23\textwidth}
      \centering
      \includegraphics[width=\textwidth]{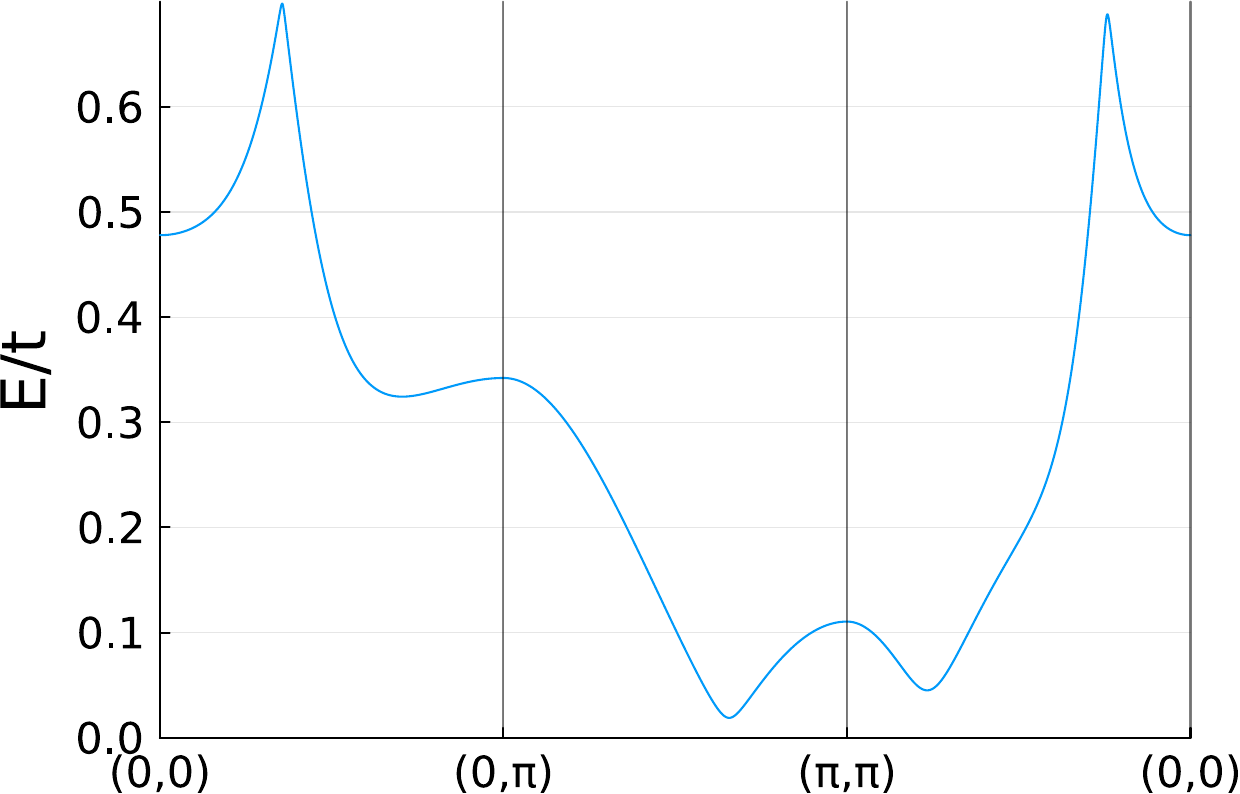}
      \caption{}
      \label{fig:5-b}
  \end{subfigure}
    % \hfill
  \begin{subfigure}[b]{0.23\textwidth}
      \centering
      \includegraphics[width=\textwidth]{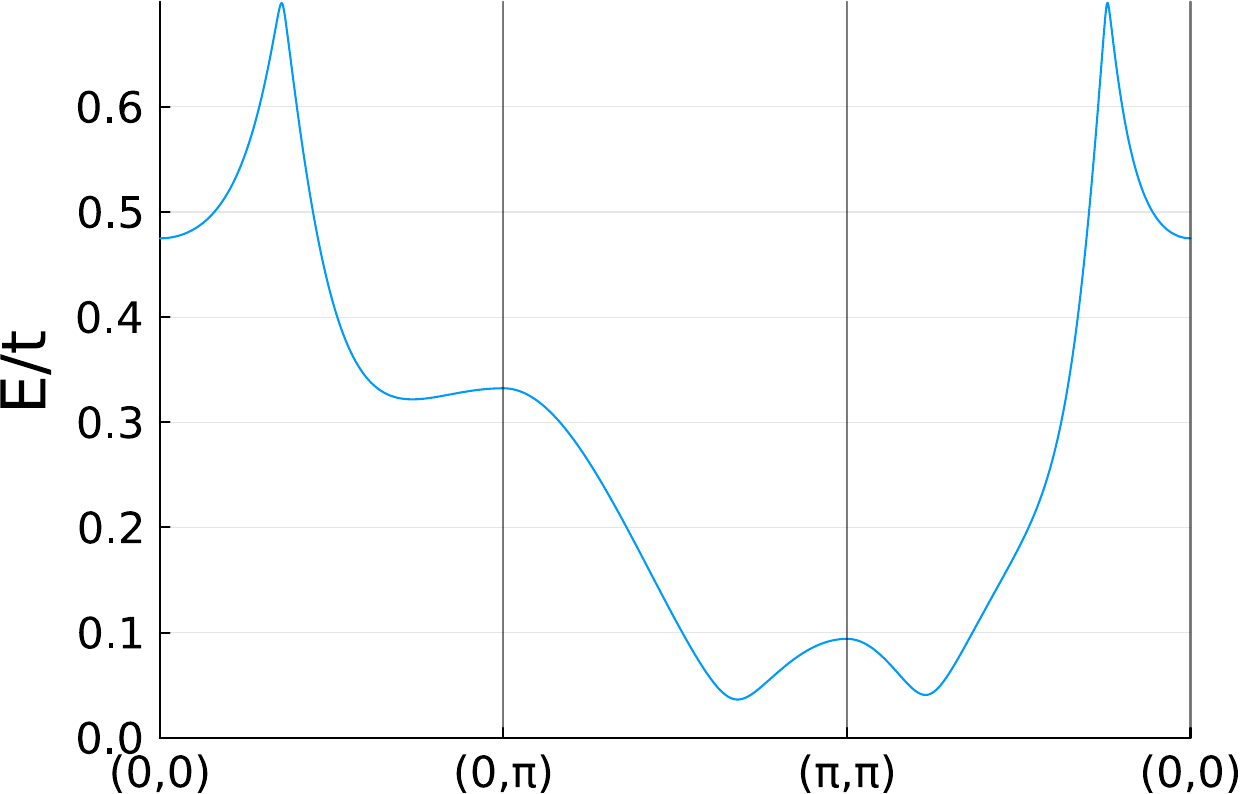}
      \caption{}
      \label{fig:5-b}
  \end{subfigure}
  \caption{Bogoliubov quasiparticle dispersions (the lowest energy band with $E\geq0$) for different values of $J_2/J_1$ at $J_K=5.8, n_{\mathrm{c}} = 0.3$.
  (a) $d_{xy}$-wave superconductivity with $\theta=0.7$,
  (b) $p+d$-wave superconductivity with $\theta=0.73$,
  (c) $p$-wave helical superconductivity with $\theta=0.76$.
  There are Dirac points on the line $k_y=\pi$ and other symmetric points in (a). 
  The quasiparticle dispersions are fully gapped in (b) and (c), where there are non-zero $p$-wave components.}
  \label{fig:5}
\end{figure}

Behaviors of the superconducting pairing amplitudes with respect to $J_2/J_1$ shown in
Fig.~\ref{fig:4}  provide insight into the emergence of the coexisting superconducting phase.
The two pairing amplitudes $\Delta^{\mathrm{sin}}$ and $\Delta^{\mathrm{tri}}$ exhibit distinct $J_2/J_1$-dependence;
the former is square root %(or possibly weakly discontinuous) 
in the parameter, 
$\Delta^{\mathrm{sin}}\sim \sqrt{J_2/J_1-(J_2/J_1)_{c1}}$, or possibly weakly discontinuous.
On the other hand, the latter shows linear dependence around the critical point, 
$\Delta^{\mathrm{tri}}\sim |J_2/J_1-(J_2/J_1)_{c2}|$.
These dependence on $J_2/J_1$ can be understood based on the Bogoliubov quasiparticle spectra.
The $d_{xy}$-wave superconducting state with the pairing amplitude $\Delta^{\mathrm{sin}}\sim \sin k_x\sin k_y$
has nodes on $k_x=0,\pi$ and $k_y=0,\pi$, and consequently the Bogoliubov quasiparticles have Dirac-like
linear dispersions around intersections of the Fermi surface and the nodes (Fig.~\ref{fig:5-a}). 
In contrast, the $p$-wave helical pairing state with $d_{\mu}\sim \sin k_{\mu}$ does not have nodes, 
and the resulting Bogoliubov quasiparticles are fully gapped (Fig.~\ref{fig:5-b}).
When $J_2$ is increased from a small value (or equivalently $J_1$ is decreased), 
there is no low energy fermionic excitations due to $\Delta^{\mathrm{tri}}\neq0$ and
a conventional bosonic phase transition for the order parameter $\Delta^{\mathrm{sin}}$ 
is driven by the interaction $J_2$, resulting in the 
standard square root behavior within the mean field approximation.
The true critical behavior beyond the mean field approximation will belong to the XY universality class 
in (2+1)-dimension, if the quantum phase transition is continuous. 
On the other hand, when  $J_2$ is decreased from a large value (or equivalently $J_1$ is increased),
there are gapless Dirac fermion excitations due to the nodes of
$\Delta^{\mathrm{sin}}$ and the bosonic field $\Delta^{\mathrm{tri}}$ interact with the Dirac fermion excitations.
As a result, the corresponding quantum phase transition driven by the $J_1$ interaction is not
a conventional bosonic one, but is a ``fermionic" quantum phase transition.
The linear dependence of the order parameter within the mean field approximation is 
a characteristic feature of a qunatum phase transition in an interacting 
Dirac system~\cite{Sorella1992-op, Rosenstein1993, Herbut2006, Wang2014, Ohtsuka2016,Tada2020}.
Beyond the mean field approximation, 
the true criticality will be the (2+1)-dimensional chiral XY universality class with eight
Dirac points in the Brillouin zone.

%%%%%%%%%%%%%%%%%%%%%%%%%%%%%%%%
\subsection{High conduction electron filling ($n_c=0.8$)}

\begin{figure}[htb]
  \centering
  \includegraphics[width=8cm]{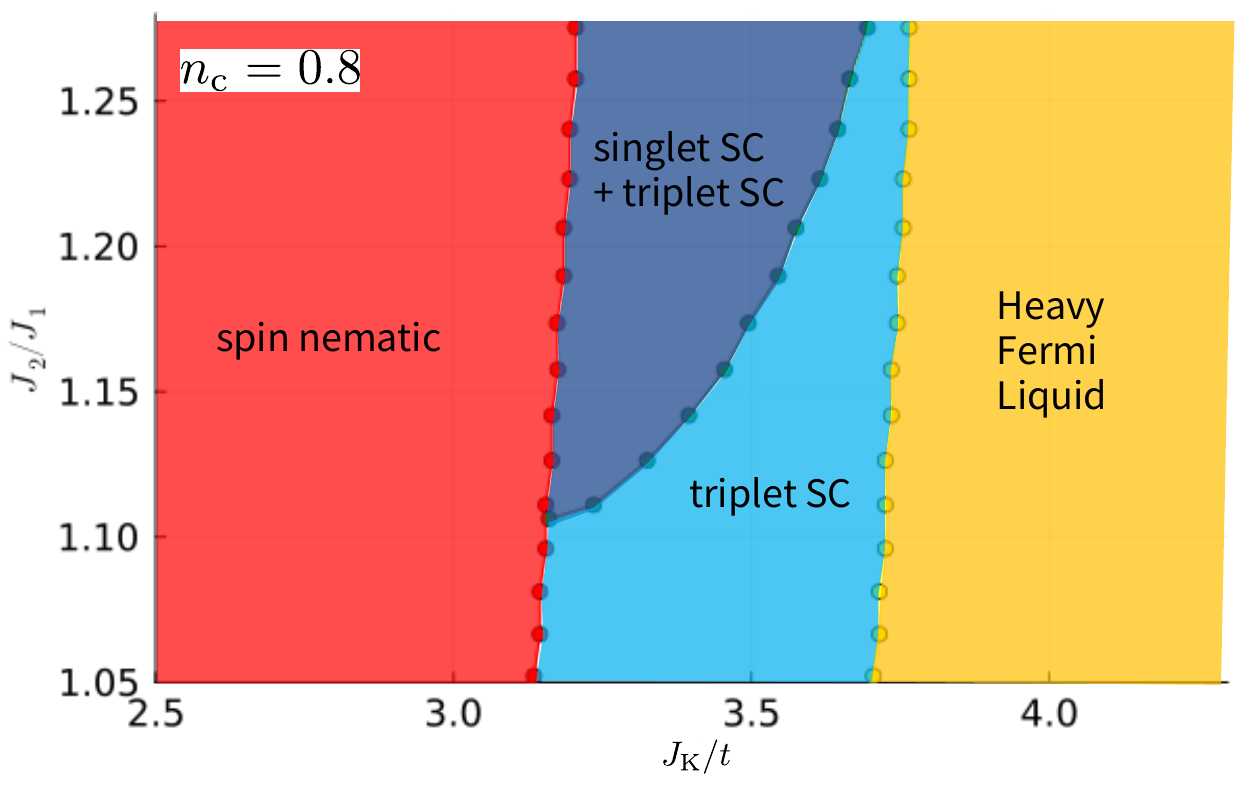}% Here is how to import EPS art
  \caption{The phase diagram of the $J_1$-$J_2$ Kondo lattice model at $n_\mathrm{c}=0.8$.
In the red region,  the spin nematic state is stable (i.e. FL$^*$). The blue regions are the superconducting states, and the
yellow region is the HFL.
In the triplet superconducting state, the helical pairing and the chiral pairing are degenerate in the mean field approximation.}
  \label{fig:7}
\end{figure}

Next, we briefly discuss the high filling ($n_c=0.8$) case. 
In Fig.~\ref{fig:7}, the phase diagram at $n_c=0.8$ is presented.
First of all, the overall structure of the phase diagram at $n_c=0.8$ 
is similar to that of Fig.~\ref{fig:3} at the low filling $n_c=0.3$ in the previous section.
Namely, there is an FL$^*$ for small $J_K$, a standard HFL for large $J_K$, and superconductivity for an intermediate
$J_K$. 
Quantitatively, the FL$^*$ region at $n_c=0.8$ is stable for $J_K\lesssim 3$
and it is slightly narrower than that in the low filling case.
This would be because the density of states of the conduction electrons is relatively large at high filling
and thus the Kondo effect is stronger than that in the low filling case.
This also leads to the HFL state which is stable for a wide range of the Kondo coupling 
and the superconducting region which is rather narrow.

In the superconducting region, the mixed pairing of the helical $p$-wave pairing and the $d_{xy}$-wave pairing 
is stabilized for large $J_2/J_1$, and there is no purely spin-singlet superconductivity in contrast to the low filling case.
For relatively small $J_2/J_1$, the helical triplet pairing state and the chiral triplet pairing state
are degenerate in the pure triplet superconducting region within the mean field approximation.
An interesting difference between the low filling case and the high filling case is that 
the pure spin-triplet superconducting region is enlarged in the latter.

Spin-triplet superconductivity is often related with ferromagnetic spin correlations and 
it can be stable within mean field approximations in a model with ferromagnetic interactions. 
In the standard Kondo lattice, the ferromagnetism (antiferromagnetism) is realized in a low (high) filling region,
and the ferromagnetic order implies that triplet pairing driven by the Kondo interaction $J_K$ 
may be favored at low filling.
However,
in the present $J_1$-$J_2$ Kondo lattice, the flat-band state corresponding to the ferromagnetism 
is not stable for $J_K\neq0$ as mentioned previously, and
the enhanced triplet pairing at the high filling is
in sharp contrast to the naive expectation for the standard Kondo lattice.
The triplet superconductivity in the $J_1$-$J_2$ Kondo lattice is not mediated by the Kondo coupling $J_K$ (or RKKY interaction)
but by the ferromagnetic interaction $J_1$, 
and this mechanism is strengthened at high filling.
Similarly, as mentioned above, 
the pure spin-singlet superconductivity is not found in the present parameter region,
although the singlet pairing is often related to antiferromagnetism.
The singlet pairing is induced by the antiferromagnetic $J_2$ interaction and this mechanism is suppressed at high filling.
In this way, the superconductivity depends on details of the underlying large Fermi surface with both the conduction
electrons and the localized spins,
which is non-trivial interplay between the frustration and the Kondo effect.

%%%%%%%%%%%%%%%%%%%%%%%%%%%%%%%%%%%%%%%%%%%%%%%%%%%%%%%%%%%%%%
\section{\label{sec:level4}summary}
In this work, we have studied competition and cooperation of the frustration and the Kondo effect 
in the $J_1$-$J_2$ Kondo lattice in two dimensions within the slave boson mean field approximation.
In absence of the Kondo coupling $J_K$, the localized spins show the spin nematic order which could be 
regarded as a variant of spin liquids with fractionalized spinon excitations as discussed in the previous study.
We found that the small non-zero Kondo coupling does not destroy the spin nematic order 
and the Kondo effect is absent. The localized spins and the conduction electrons are effectively
decoupled, and the ground state is regarded as an FL$^*$ state with a small Fermi surface.
For sufficiently large $J_K$, the spin nematic order is completely suppressed and the Kondo effect
leads to the formation of the standard HFL with a large Fermi surface.
In the intermediate region of $J_K$, the superconducting states can arise from the interplay of 
the spin nematic order and the Kondo effect.
It is found that the spin-singlet pairing has $d_{xy}$-wave symmetry and the spin-triplet pairing is
the time-reversal symmetric helical $p$-wave state when the singlet-triplet mixture takes place.
In the pure triplet state, the helical state and the time-reversal symmetry broken chiral $p$-wave 
state are degenerate within the mean field approximation. 
We also discussed the nature of the quantum phase transitions of the superconductivity,
and found the distinct critical behaviors of the spin-singlet component and the spin-triplet one belonging to 
the different universality classes associated to the absence or existence of the gapless Dirac fermion excitations.
The stability of the superconductivity depends on the conduction electron filling in a non-trivial way.

In the present study, we have focused on the spin nematic order at zero magnetic field within 
the slave boson mean field approximation.
Although the $J_1$-$J_2$ model is expected to be located close to the zero field spin nematic order,
non-zero magnetic fields stabilize the spin nematic state which is described by magnon pair condensation.
It would be an interesting future study to investigate the spin nematic order coupled to the conduction electrons
under magnetic fields.
Another possible direction is to examine a Kondo lattice model with classical spins
where coupling with conduction electrons could lead to various magnetic orders
~\cite{Shannon2010,Akagi2010,Reja2015}.

%%%%%%%%%%%%%%%%%%%%%%%%%%%%%%%%%%%%%
\appendix
\section{Gauge fluctuations and SU(2) formulation}
\label{app:su2}
%%%%%%%%%%%%%%%%%%%%%%%%%%%%%%%%%%%
We briefly discuss gauge fluctuations in the $J_1$-$J_2$ Kondo lattice model
based on the SU(2) formulation of the $J_1$-$J_2$ spin model proposed in the previous study~\cite{Shindou2009-fj}.
The SU(2) formulation is convenient for a discussion of gauge fields~\cite{Wen2007,Wen1991,Wen2002,Sachdev2023}.
In the path integral formulation, the Lagrangian contains spinon-boson coupling terms
$(J_1/4){\rm tr}[F^{\dagger}_iU_{ij}^{\rm tri}F_j\sigma_{\mu}^T]$ and $(J_2/4){\rm tr}[F^{\dagger}_iU_{ij}^{\rm sin}F_j]$,
where $F_i$ is the fermionic spinon field and $U_{ij}^{\rm sin},U_{ij,\mu}^{\rm tri}$ are the bosonic Hubbard-Stratonovich field corresponding to the decoupling Eqs.~\eqref{eq:dec1} and ~\eqref{eq:dec2},
%%%%%%%%%%%%%%
\begin{align}
F_j&=\left(
\begin{array}{cc}
f_{j\uparrow} & f_{j\downarrow} \\
f_{j\downarrow}^{\dagger} & -f_{j\uparrow}^{\dagger}
\end{array}
\right), \\
U^{\rm sin}_{jk}&=\left(
\begin{array}{cc}
\chi_{jk}^{\ast} & \eta_{jk} \\
\eta_{jk}^{\ast} & -\chi_{jk}
\end{array}
\right), \quad
U^{\rm tri}_{jk,\mu}&=\left(
\begin{array}{cc}
E_{jk,\mu}^{\ast} & D_{jk,\mu} \\
-D_{jk,\mu}^{\ast} & E_{jk,\mu}
\end{array}
\right).
\end{align}
%%%%%%%%%%%%%%
These are all dynamical fields to be integrated in the path integral formulation.
The action $S_{J_1\mbox {-}J_2}[F,U]$ of the $J_1$-$J_2$ Hamiltonian has the gauge redundancy,
%%%%%%%%%%%%%%
\begin{align}
F_j\to g_j F_j, \quad
U_{jk}^{\rm sin/tri}\to g_jU_{jk}^{\rm sin/tri}g_k^{-1},
\end{align}
%%%%%%%%%%%%%%
where $g_j$ is a local SU(2) matrix at site $j$.
Corresponding to the decoupling Eq.~\eqref{eq:decK},
we can introduce a similar matrix fields
for the conduction electrons and the Hubbard-Stratonovich field $\hat{V}_j$ so that the Kondo coupling is rewritten as 
$J_K\sum_i {\rm Tr}[ C_j^{\dagger}\hat{V}_jF_j]$,
%%%%%%%%%%%%%%
\begin{align}
C_j&=\left(
\begin{array}{cc}
c_{j\uparrow} & c_{j\downarrow} \\
c_{j\downarrow}^{\dagger} & -c_{j\uparrow}^{\dagger}
\end{array}
\right), \quad
\hat{V}_j=\left(
\begin{array}{cc}
V_j & 0 \\
0 & -V_j^{\dagger}
\end{array}
\right).
\end{align}
%%%%%%%%%%%%%%
The Kondo coupling term $S_{K}[F,C,V]$ is invariant under the local SU(2) transformation
%%%%%%%%%%%%%%
\begin{align}
F_j\to g_j F_j, \quad
\hat{V}_j\to \hat{V}_jg_j^{-1}.
\end{align}
%%%%%%%%%%%%%%
Note that $\hat{V}_j$ transforms non-trivially ($\hat{V}_jg_j^{-1}\neq \hat{V}_j$) 
and it can have non-zero off-diagonal elements for a general SU(2) gauge choice~\cite{flint2008heavy}.

We point out that the boson field $\hat{V}_j$ has an SU(2) gauge charge
in addition to the standard U(1) charge of the conduction electrons, simply because it corresponds to a composite field
of the spinons and the conduction electrons, $V\sim f^{\dagger}c$.
Therefore, it can behave as Higgs fields for various gauge fluctuations as seen below.
In the saddle-point approximation, 
we simply replace the dynamical fields $U$ and $V$ by static mean fields $\bar{U}$ and $\bar{V}$.
Low energy fluctuations around the mean field $\bar{U}$ can be 
found based on the so-called invariant gauge group of the 
mean field action $S^{(\bar{U},\bar{V})}[F,C]$.
The invariant gauge group for the total system is defined by the condition
$g_j\bar{U}_{jk}g_k^{-1}=\bar{U}_{jk}$ when $\bar{V}=0$.
In the FL$^*$ phase ($\bar{V}=0$), there can be gapless gauge fluctuations as discussed previously
~\cite{Shindou2009-fj}.
For example, 
in the FL$^*$ state with the purely triplet spin nematic order ($\bar{D}\neq0,\bar{\eta}=0$), 
the gauge flux is $\bar{U}^{\rm tri}_{j,j+\hat{x},1}\bar{U}^{\rm tri}_{j+\hat{x},j+\hat{x}+\hat{y},2}\bar{U}^{\rm sin}_{j+\hat{x}+\hat{y},j}
=- \bar{D}^2\bar{\chi}\sigma_3$.
This flux induces mass terms for the gauge fields $a_1\sigma_1, a_2\sigma_2$ at low energy, and 
the SU(2) gauge fluctuations are effectively reduced to those of the remaining gapless U(1) gauge field $a_3\sigma_3$
around the mean field ansatz.
Indeed, the saddle point ansatz $\bar{U}$ in Eq.~\eqref{eq:helical_ansatz} has the invariant gauge group symmetry
$g_j=e^{i(-1)^{j}\theta\sigma_3}, \bar{U}_{jk}=g_j\bar{U}_{jk}g_k^{-1}$ ($\theta$ is a constant), which protects the U(1) gauge fluctuation near
the $k$-point $(\pi,\pi)$ in the Brillouin zone. 
The monopole fluctuations 
in the pure compact U(1) gauge theory in $2+1$ dimensions lead to the confinement of the spinon gauge charge
and the mean field ansatz of the pure triplet spin nematic order is unstable.
On the other hand, 
the invariant gauge group does not contain the above U(1) transformation once
Higgs condensation ($\bar{V}\neq 0$) takes place.
This means that a mass term of the U(1) gauge field is generated by the non-zero Higgs condensation
in an effective action of the gauge field.
Therefore,
the helical superconducting state is stabilized by the Kondo effect as discussed in the main text. 
Similarly, the Kondo coupling term at the saddle-point is not invariant for other gauge transformations such as 
$g_j=e^{i\theta \sigma_2}, e^{i\theta \sigma_3}$.
In this way, the invariant gauge groups of the total saddle-point action $S^{(\bar{U},\bar{V})}[F,C]$ for 
the superconducting states and the HFL state with $\bar{V}\neq 0$ 
do not have continuous transformation
and correspondingly there are no gapless gauge fluctuations in these phases.

\begin{acknowledgments}
We thank Han Yan for helpful discussions.
This work is supported by JSPS KAKENHI Grant No. 22K03513.
\end{acknowledgments}

%\nocite{*}
\bibliography{reference}% Produces the bibliography via BibTeX.
\end{document}